# Fractal Analytical Approach of Urban Form Based on Spatial Correlation Function

Yanguang Chen

(Department of Geography, College of Urban and Environmental Sciences, Peking University, 100871, Beijing, China. Email: chenyg@pku.edu.cn )

**Abstract:** Urban form has been empirically demonstrated to be of scaling invariance and can be described with fractal geometry. However, the rational range of fractal dimension value and the relationships between various fractal indicators of cities are not yet revealed in theory. By mathematical deduction and transformation (e.g. Fourier transform), I find that scaling analysis, spectral analysis, and spatial correlation analysis are all associated with fractal concepts and can be integrated into a new approach to fractal analysis of cities. This method can be termed '3S analyses' of urban form. Using the 3S analysis, I derived a set of fractal parameter equations, by which different fractal parameters of cities can be linked up with one another. Each fractal parameter has its own reasonable extent of values. According to the fractal parameter equations, the intersection of the rational ranges of different fractal parameters suggests the proper scale of the fractal dimension of urban patterns, which varies from 1.5 to 2. The fractal dimension equations based on the 3S analysis and the numerical relationships between different fractal parameters are useful for geographers to understand urban evolution and potentially helpful for future city planning.

# 1 Introduction

Fractal geometry is a powerful tool in geographical modeling and spatial analysis, and it has been applied to urban studies for a long time (Arlinghaus, 1985; Arlinghaus and Arlinghaus, 1989; Batty, 1995; Batty and Longley, 1994; Benguigui and Daoud, 1991; De Keersmaecker *et al*, 2003;

Frankhauser, 1998; Thomas et al, 2010). The fractal concept is in essence based on scaling symmetry, and symmetry suggests invariance under some kind of transformation (Mandelbrot, 1989). Thus self-similarity is equivalent to invariance under contraction or dilation, which is termed 'scaling invariance'. In mathematics, fractal property can be abstracted as scaling relations between linear scales and corresponding measurements. A city can be empirically treated as a fractal system with self-similarity or self-affinity (e.g. Batty, 2008; Batty, 2005; Benguigui et al, 2000; Chen, 2010; Feng and Chen, 2010; Frankhauser, 1994; Thomas et al, 2007; Thomas and Frankhauser et al, 2008; White and Engelen, 1993; White and Engelen, 1994). Where urban form is concerned, the relation between the radius (a scale) from a city center and the corresponding urban density (a measurement) may follow the inverse power law indicating a fractal distribution (Batty and Longley, 1994; Batty and Xie, 1999; Frankhauser, 1994). Smeed's model on traffic network can be employed to analyze urban growth and estimate fractal dimension of urban patterns (Batty and Longley, 1994).

The spatial structure of fractal cities follows power laws or inverse power laws, which implies some kind of scaling relations. Smeed's model of urban density suggests an inverse power-law distribution (Smeed, 1963). A power-law distribution is a scale-free distribution that cannot be effectively described with the conventional statistical measures such as mean, variance, and covariance. The scale-free distributions can be dealt with by scaling analysis in both theoretical and empirical studies. In theory, scaling analysis is an approach to deriving a particular power-law relation from a certain equation. The derivation process is always based on the property of invariance under scaling transform (contraction/dilation transform). From the power-law relation we can obtain one or more useful parameters, which are termed 'scaling exponents' and usually associated with fractal dimension. In practice, scaling analysis indicates an empirical separation process of a system, say, a city, into different aspects by means of scaling exponents (Jiang and Yao, 2010). The scaling analysis is very useful and significant in both theoretical and empirical studies of urban patterns and processes.

The inverse power function of urban density mentioned above proved to be a special spatial correlation function (Takayasu, 1990). Urban growth and form have been modeled by using the concept of spatial correlation (Makse et al, 1995; Makse et al, 1998). Spatial correlation is an underlying way of modeling both urban growth and form, and the correlation models can be mathematically analyzed by scaling transform. In fact, based on the density function, a general spatial correlation function of cities can be constructed (Chen, 2011; Chen and Jiang, 2010). If the correlation function is converted into energy spectrum, the urban density will be converted into spectral density through Fourier transform (Chen, 2008). Thus spatial correlation analysis can be converted into spectral analysis and *vice versa* (Chen, 2009). Both spatial correlation analysis and spectral analysis are useful tools in urban studies (Cauvin *et al*, 1985; Chen and Jiang, 2010). The two analytical processes are associated with scaling analysis. A problem is how to combine the scaling analysis, spectral analysis, and spatial correlation analysis with one another to form a new method of spatial analysis of cities.

Especially, in order to apply fractal theory to city planning and urban spatial optimization, the geographical meaning of fractal dimension values must be revealed. In the previous works, the fractal dimension values of urban patterns were discussed by Frankhauser (2004), and the statistical relationship between residential satisfaction and the fractal dimension of the built-up residential environments was discussed by Thomas and Tannier *et al* (2008). This paper is devoted to revealing the theoretical relations and proper numerical scales of different fractal parameters of urban form. Based on the inverse power law of urban density, scaling analysis, spectral analysis, and spatial correlation analysis will be integrated to make a new approach to analyzing urban patterns and process (Section 2). As a case study, the methods will be applied to three cities in Yangtze River delta, China (Section 3). The academic contributions of this paper to fractal theory of cities are as follows. First, a new method termed '3S analysis' of fractal city systems is presented, and the analytical procedure is sketched out. Second, a set of fractal parameter equations is derived, and these equations are useful for our understanding urban development. Third, the valid range of fractal indicators of urban form is determined, and the results are potentially helpful for future city planning.

## 2 Theoretical argument

### 2.1 Type of data used

A fractal city is generally defined in a 2-dimension space based on a digital map or a remotely sensed image (Batty and Longley, 1994; Frankhauser, 1994). In other words, the dimension of the embedding space of a city fractal is $d=2$. The fractal dimension values of urban patterns can be estimated with the box-counting methods (Benguigui *et al*, 2000; Chen, 2012a; Feng and Chen, 2010; Shen, 2002), the area-radius scaling (Batty and Longley, 1994; Chen, 2010; Frankhauser, 1994), the area-perimeter scaling (Batty and Longley, 1994; Wang *et al*, 2005), and so on. Each method has its strong point. If we want to examine the patterns of spatial distribution of urban built-up elements, the box-counting method is the best approach to evaluating fractal dimension. However, if we want to investigate the process of urban growth, the area-radius scaling is the best way of estimating fractal dimension because this manner of procedure is more consistent with the relation between urban core and periphery than other methods.

Suppose that the fractal dimension of urban form will be evaluated by the remote sensing data of a city from the angle of view of urban growth. We can compute the urban area or the number of cells (the smallest image-forming unit of a digital map, i.e., pixels), $N(r)$, within the radius, $r$, from the city center. If the relation between measurement $N(r)$ and the scale $r$ follow a power law, then the urban form defined on a digital map can be regarded as a fractal pattern, and the scaling exponent is just the fractal dimension termed *radial dimension* (Frankhauser and Sadler, 1991). Equivalently, we can investigate the relation between the radius $r$ from a city center and the corresponding urban density $\rho(r)$. If the relation follow a scaling law, and if the value of the power exponent comes between 0 and 1, then the urban growth can be treated as a fractal process with a fractional dimension value varying from 0 to 2 (Chen, 2010).

This study is based on density-radius scaling relation of urban form. The concepts of urban form and urban density should be clarified before the theoretical models are presented. Based on a 2-dimensional space, *urban form* can be defined as the spatial pattern of elements composing the city in terms of its networks, buildings and spaces (Batty and Longley, 1994). Thus, *urban density* refers to the number of inhabitants, buildings, roads/streets, in given urbanized area. The urban density has different connotations for different spatial measurements. If we study urban population distribution, the urban density indicates urban population density; if we research the patterns of urban land uses, the urban density implies urban land-use density; if we investigate the spatial structure of transport network, the urban density suggests the urban road density.

## 2.2 The correlation dimensions

The theoretical starting point of this work is monocentric cities, but possibly the conclusions in this paper can be generalized to polycentric cities. For a city of self-similarity with a growth core, the urban density $\rho(r)$ at distance $r$ from the urban center is

$$\rho(r) = \rho_1 r^{D_f - d} = \rho_1 r^{-a}, \qquad (1)$$

where $\rho_1$ is a proportionality coefficient, $a=d-D_f$ denotes the scaling exponent of density distribution, $d$ refers to Euclidean dimension, and $D_f$, to the *radial dimension* of urban form ($D_f<d$) (Frankhauser and Sadler, 1991). Equation (1) is actually Smeed's model (Smeed, 1963; Batty and Longley, 1994). Because there is no mathematical definition of the city center, we can specially define a central density $\rho(0)=\rho_0$ for the point $r=0$. Even if the urban density function is defined in a 1-dimension space based on the concept of statistical average, it reflects the system information in a 2-dimension space. Generally speaking, the fractal dimension of urban form ($D_f$) comes between 1 and 2, and the scaling exponent of urban density ($a$) ranges from 0 to 1. In practice, sometimes $D_f<1$ or even $D_f>2$, and thus we have $a>1$ or $a<0$.

Fractal analysis is often related with correlation analysis because a fractal model is usually associated with a correlation function. The generalized fractal dimension is what is called correlation dimension (Chen and Jiang, 2010; Grassberger and Procaccia, 1983). Now, let's consider two points on a radial from the city center, X and Y (Figure 1a). The density-density correlation function based on equation (1) can be constructed as

$$C(r) = \int_{-\infty}^{\infty} \rho(x)\rho(x+r)dx = 2\rho_1^2 \int_0^{\infty} x^{D_f - d}(x+r)^{D_f - d} dx, \qquad (2)$$

in which $x$ refers to the distance of the first point (X) from the city center, and $r$ to the distance of the second point (Y) from the first point (X). Note that the density function is not continuous at $x=0$. It is easy to prove that the spatial correlation function follows the scaling law under dilation. Given a function $f(x)$, if our scaling the argument $x$ by a constant scale factor $\lambda$ causes only a proportionate scaling of the function itself, that is $f(\lambda x)= \lambda^\alpha f(x)$, where $\alpha$ is a scaling exponent, then we will say that the function complies with the scaling law. Let $x=\xi y$, in which $\xi$ is a scale factor. A scaling relation can be demonstrated in the form

$$\begin{aligned}
C(\xi r) &= 2\rho_1^2 \int_0^\infty x^{D_f - d}(x + \xi r)^{D_f - d} dx \\
&= 2\rho_1^2 \int_0^\infty (\xi y)^{D_f - d}(\xi y + \xi r)^{D_f - d} d(\xi y) \\
&= \xi^{2(D_f - d) + 1} 2\rho_1^2 \int_0^\infty y^{D_f - d}(y + r)^{D_f - d} dy \\
&= \xi^{2H} C(r)
\end{aligned} \quad (3)$$

where $2H = 2(D_f - d) + 1$, and $H$ proved to be the generalized Hurst exponent. In the theory of R/S analysis, the Hurst exponent is a scaling exponent associated with the autocorrelations of the time series (Feder, 1988; Hurst *et al*, 1965). The R/S analysis is generally termed 'rescaled range analysis'. This is a statistical method developed by Hurst *et al* (1965) to analyze long-term continuous or regular records of natural phenomena. It can also be employed to analyze the long orderly spatial measurements with proportional spacing. The rescaled range is a statistical measure of the variability of a time/space series based on two main measures/variables: one is the standard deviation ($S$), and the other, the range ($R$) of the data set (the difference between the highest and lowest values). From the slope of the logarithmic linear relation between the ratio of $R(\tau)$ to $S(\tau)$ and the time/space lag $\tau$, we can obtain a useful parameters, the Hurst exponent ($H$). Concretely speaking, for the increment series $\Delta x(i)$ of a space/time series $x(i)$, $H$ is the scaling exponent of the ratio $R(\tau)/S(\tau)$ versus time/space lag ($\tau$) ($i=1,2,3,\ldots$; $\tau=1,2,\ldots,i$). In other words, $H$ is defined by the power function $R(\tau)/S(\tau) = (\tau/2)^H$ (Chen, 2010; Feder, 1988). The exponent value comes between 0 and 1 ($0 \le H \le 1$). The value of $H = 1/2$ indicates a Brownian motion, while the values of $H \ne 1/2$ suggests the fractional Brownian motion (fBm) (Feder, 1988; Mandelbrot, 1983). Apparently, the solution to the above functional equation is

$$C(r) = C_1 r^{2(D_f - d) + 1} = C_1 r^{2H} = C_1 r^b, \quad (4)$$

in which $C_1$ refers to a constant of proportionality, and $b = 2H$ is a scaling exponent.

The correlation function, equation (4), is defined in a 1-dimension space, so it differs from the spatial correlation defined in a 2-dimension space (Chen and Jiang, 2010). For the latter, the scaling exponent is just the correlation dimension, but for the former, the scaling exponent is less than the correlation dimension. By dimensional analysis, we have

$$N(r) \propto rC(r) = r^{2 - 2(d - D_f)} = r^{2(D_f - 1)} = r^{D_c}, \quad (5)$$

where $N(r)$ denotes the number of pixels within the radius of $r$ from the center, and $D_c$ the density-density correlation dimension, or the *point-point correlation dimension* of urban density. Equation (5) gives such a dimension relation

$$D_c = 2(D_f - 1).  \qquad (6)$$

For the monocentric cities, the radial dimension can be directly calculated by the area-radius scaling (Batty and Longley, 1994; Frankhauser, 1998). However, for the polycentric cities, the area-radius scaling relations often break down and the radial dimension cannot be well estimated in a simple way. In this instance, the spectral analysis is necessary due to the filter function of Fourier transform (Chen, 2010).

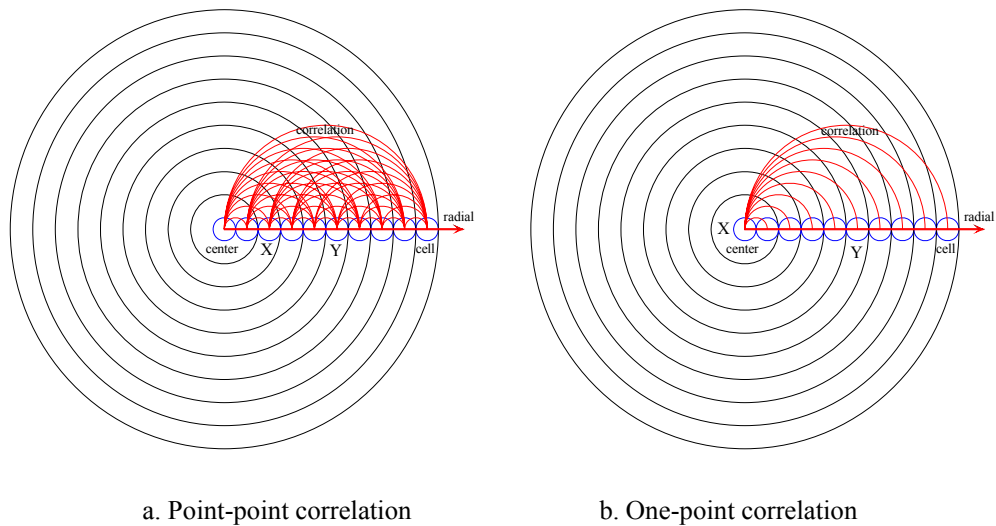

a. Point-point correlation    b. One-point correlation

**Figure 1 The sketch maps of spatial correlation based on the concept of statistical average**

**Note:** The systems of concentric circles are used to compute average density. The space between two immediate circles forms a ring. The little circles within the rings represent cells along a radial, and the arcs denote spatial correlation among cells.

One of the special cases of the urban density-density correlation is the central correlation. It can be treated as a kind of "one-point correlation". The one-point correlation indicates the spatial correlation between a given point and other points around the point, while the point-point correlation implies the spatial correlation between one point on a circle and another point on another circle around the center of a city. Suppose that there is a radial from the city center to outside (circumference). The density-density correlation indicates the spatial relation between any two points on the radial. If we fix one point, say, X, to the center of city, we will have $x=0$ (Figure 1b). Thus the point-point correlation function, equation (2), is reduced to a special one-point correlation function such as

$$C_0(r) = \rho_0 \rho_1 r^{D_f - d} = \rho_0 \rho(r), \qquad (7)$$

where $C_0(r)$ denotes the one-point correlation measurement. This suggests that, if we fix one point in the city center, the one-point correlation function is proportional to the urban density function, thus the correlation integral is in (direct) proportion to the pixel number within the radius of $r$ from the center, $N(r)$. The radial dimension of cities is just the one-point correlation dimension, i.e., $D_0 = D_f$, where $D_0$ denotes the *central* or *one-point correlation dimension*. Therefore, $a = d - D_f = d - D_0$ indicates the scaling exponent of the one-point correlation, which suggests the spatial relation and interaction between the core and the periphery of a city.

A correlation function can be converted into energy spectrum through Fourier transform and *vice versa*. The Fourier transform of the density-density correlation function also follows the scaling law, and this can be proved as below

$$\begin{aligned} S(\xi k) &= C_1 \int_{-\infty}^{\infty} r^{1-2(d-D_f)} e^{-i 2\pi \xi k r} \mathrm{d}r \\ &= \xi^{-2+2(d-D_f)} C_1 \int_{-\infty}^{\infty} (\xi r)^{1-2(d-D_f)} e^{-i 2\pi k (\xi r)} \mathrm{d}(\xi r) \\ &= \xi^{-2(D_f - 1)} S(k) \end{aligned} \qquad (8)$$

where $k$ denotes the wave number. Thus we have

$$S(k) \propto k^{-2(D_f - 1)} = k^{-\beta}, \qquad (9)$$

where $\beta = 2(D_f - 1) = 2H + 1$ refers to the spectral exponent. This suggests that the spectral exponent equals the density-density correlation dimension, that is

$$\beta = D_c = 2(D_f - 1), \qquad (10)$$

which has been empirically confirmed by Chen (2010), who applied it to Beijing city, China. Comparing equation (5) with equation (9) yields the relation between the spectral density and the correlation function

$$N(r) \propto r C(r) \propto \frac{1}{S(k)}. \qquad (11)$$

This implies that the cell/pixel number of urban land use varies inversely as the spectral density.

For the central correlation, the cosine transform relation between the correlation function and energy spectrum cannot hold. The Fourier transform of the one-point correlation function yields

$$F(k) = \int_{-\infty}^{\infty} C_0(r) e^{-i2\pi kr} \mathrm{d}r = \rho_0 \rho_1 \int_{-\infty}^{\infty} r^{D_0-d} e^{-i2\pi kr} \mathrm{d}r, \tag{12}$$

which proved to follow the scaling law under dilation, that is

$$\begin{aligned}F(\xi k) &= \rho_0 \rho_1 \int_{-\infty}^{\infty} r^{D_0-d} e^{-i2\pi\xi kr} \mathrm{d}r \\ &= \xi^{-1-(D_0-d)} \rho_0 \rho_1 \int_{-\infty}^{\infty} (\xi r)^{D_0-d} e^{-i2\pi k(\xi r)} \mathrm{d}(\xi r) \\ &= \xi^{-(D_0-d+1)} F(k)\end{aligned} \tag{13}$$

This implies that the energy spectrum also satisfies the scaling relation

$$S_0(\xi k) = |F(\xi k)|^2 = \xi^{-2(D_0-d+1)} |F(k)|^2 = \xi^{-2(D_0-1)} S_0(k). \tag{14}$$

The solution to equation (14) is

$$S_0(k) \propto k^{-2(D_0-1)} = k^{-\beta_0}. \tag{15}$$

Obviously, the spectral exponent of the central correlation equals that of the density-density correlation, namely, $\beta_0=\beta$. To sum up, we have the following useful parameter relation

$$\beta = 2(D_f - 1) = D_c = 2(D_0 - 1) = \beta_0. \tag{16}$$

Based on equation (16), more dimension equations can be derived for spatial analysis of cities.

## 2.3 Fractal parameter equations

The fractal parameters based on wave-spectrum scaling are defined at the macro level of urban form. Now, let's turn to the micro level to investigate the spatial autocorrelation of urban growth. The macro level is based on the density function, while the micro level is based on the density increment function. An integral of equation (1) in the 2-dimension space yields the area-radius scaling relation (Batty and Longley, 1994)

$$N(r) = N_1 r^{D_f}, \tag{17}$$

where $N_1$ is a proportionality constant. An average density formula can be derived from equation (17), that is

$$\rho^*(r) = \frac{N(r)}{A(r)} = \frac{N_1}{\pi} r^{D_f-2} \propto \rho(r), \tag{18}$$

where $\rho^*(r)$ denotes the average density within a radius of $r$ of a city center, and $A(r)=\pi r^2$ is the area of the circular field with the radius $r$. This implies that the average density $\rho^*(r)$ is proportional to the marginal density $\rho(r)$. Further, consider the variance of the density increment.

Because of symmetry of urban density function (from $-\infty$ to 0 then to $\infty$), the mean value of the density increment can be regarded as zero. Therefore the variance can be defined as

$$V(r) = \int_{-\infty}^{\infty} [\Delta\rho(x) - 0]^2 dx = \int_{-\infty}^{\infty} [\rho(x+r) - \rho(x)]^2 dx. \quad (19)$$

Making a scaling analysis yields

$$\begin{aligned} V(\xi r) &= \rho_1^2 \int_0^{\infty} [(x+\xi r)^{D_f - 2} - x^{D_f - 2}]^2 dx \\ &= \rho_1^2 \int_0^{\infty} [(\xi y + \xi r)^{D_f - 2} - (\xi y)^{D_f - 2}]^2 d(\xi y) \\ &= \xi^{2(D_f - 2) + 1} \rho_1^2 \int_0^{\infty} [(y+r)^{D_f - 2} - y^{D_f - 2}]^2 dy \\ &= \xi^{2H} V(r) \end{aligned} \quad (20)$$

where $H = 1/2 - (d - D_f)$ is the generalized Hurst exponent. The solution to equation (20) is

$$V(r) \propto r^{2H}. \quad (21)$$

If the series of density changes is a white noise, the density increment $\Delta\rho(x) = \rho(x+r) - \rho(x)$ can be thought of as a random walk, i.e. Brownian motion. If so, we have $H = 1/2$, and this suggests $D_f = 2$. However, the urban density increment is not a white noise. Because $D_f$ value comes between 1 and 2, $H$ values should vary from -1/2 to 1/2. In this case, the spatial process is always treated as an fBm process (Feder, 1988; Peitgen et al, 2004).

Then, another fractal dimension can be derived by using dimensional analysis, which is useful in human geography (Haynes, 1975). Comparing equation (21) with equation (4) shows that the variance of density increment is directly proportional to the spatial correlation function, i.e., $V(r) \propto C(r)$. The square root of the variance is the standard deviation

$$s(r) \propto r^H. \quad (22)$$

If the radial dimension $D_f$ comes between 1.5 and 2, the scaling exponent $a$ will fall between 0 and 0.5, and thus the Hurst exponent $H = 1/2 - a$ will also come between 0 and 0.5. By dimensional consistency, we have

$$\rho(r) \propto \frac{1}{s(r)} \propto r^{-H}. \quad (23)$$

Substituting equation (23) into equation (18) yields

$$N(r) \propto \rho(r) A(r) \propto \frac{\pi r^2}{s(r)} \propto r^{2-H} = r^{D_s}. \quad (24)$$

This gives the following scaling exponent

$$D_s = 2 - H, \qquad (25)$$

which represents a fractal dimension, termed *profile dimension* of urban form (Chen, 2008). The parameter is just the self-affine record dimension of the random walk (Feder, 1988).

Summarizing the above mathematical derivation and theoretical analysis, we can obtain a series of fractal parameter equations. These relations compose a useful framework for the fractal study of urban growth and form. According to equation (3) or equation (20), for $d=2$, we have

$$D_f = H + \frac{3}{2}. \qquad (26)$$

The parameter relation, equation (26), can also be derived with dimensional analysis based on the density-density correlation function. The density function is proportional to $r^{-a}$, and squaring the function yields $r^{-2a}$. The integral of the squared density function varies directly as $r^{1-2a}$, and the second root of $r^{1-2a}$ is $r^{1/2-a}$. This suggests that $H=1/2-a=1/2-(d-D_f)=D_f-3/2$ where $d=2$ is concerned. The result is identical to those from the scaling analysis of correlation function or variance function. This indicates that the result of scaling analysis is equivalent to that of dimensional analysis.

Two important relations of fractal parameters can be derived as follows. Combining equation (25) with equation (26) yields

$$D_f + D_s = \frac{7}{2}, \qquad (27)$$

which was derived by Chen (2010). Combining equation (16) with equation (26) or equation (27) yields

$$\beta = 5 - 2D_s = 2H + 1, \qquad (28)$$

which is familiar to physicists (Feder, 1988; Takayasu, 1990). Now we have a set of fractal parameter relations comprising equation (6), equation (16), equation (25), equation (26), equation (27), and equation (28).

The equations and parameters can be logically organized into a system of fractal dimensions and relations (Figure 2). In theory, if we evaluate the radial dimension $D_f$, then we will be able to know the central correlation dimension $D_0$, the density-density correlation dimension $D_c$, the spectral exponent $\beta$, the Hurst exponent $H$, and the self-affine record dimension $D_s$. Using this systematic framework, we can reveal the numerical relations of different parameters (Table 1). If the measurement center of radial dimension is properly located on a digital map, a fractal parameter should be in scale with another one in value. The valid scale of different fractal parameters are as follows: $0 \leq D_f = D_0 \leq 2$, $0 \leq D_c = \beta \leq 2$, $0 \leq D_s \leq 2$, $0 \leq \beta \leq 3$, $0 \leq H \leq 1$. According to multifractal theory, the density-density correlation dimension must be less than then the central correlation dimension, i.e., $\beta = D_c \leq D_f = D_0$. The parameters $D_f$, $D_0$, $D_c$, and $\beta$ are defined at the macro level, while $D_s$ and $H$ are defined at the micro level. Without considering the relation between macro level and micro level of a city, the reasonable range of radial dimension will range from 1 to 2. If $D_f < 1$, then $\beta = D_c < 0$; if $D_f > 2$, then $\beta = D_c > D_f$. If macro-micro urban relation is taken into account, the reasonable range of radial dimension will be from 1.5 to 2. If $D_f < 1.5$, then $D_s > 2$ and $H < 0$, which is of illogic. To sum up, the rational scale of radial dimension comes between 1.5 and 2, i.e., $1.5 \leq D_f \leq 2$. Correspondingly, the reasonable value of profile dimension also falls in between 1.5 and 2 (Table 1).

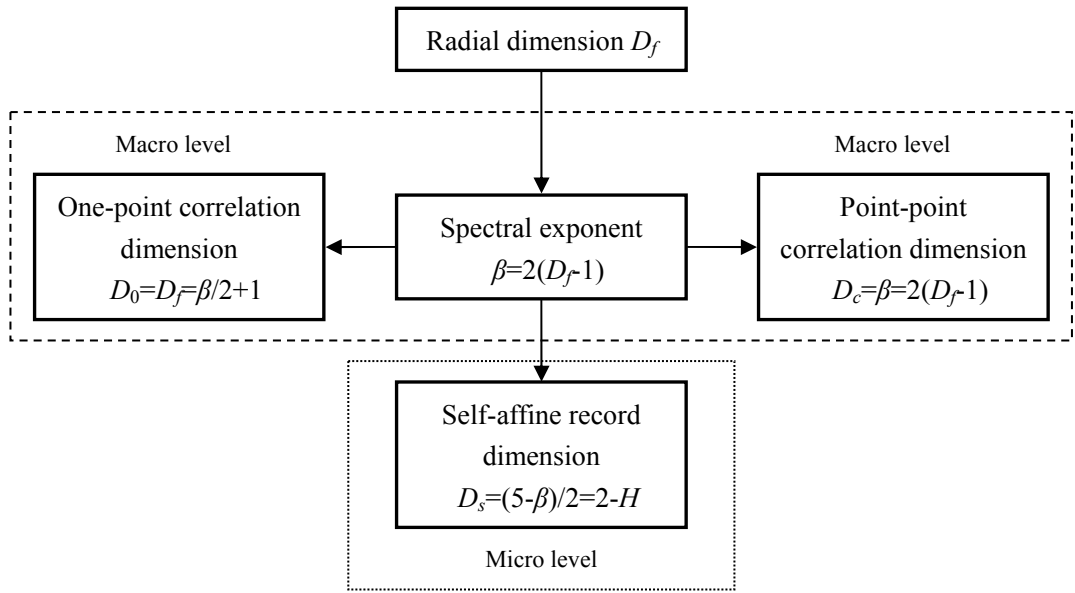

**Figure 2 A sketch map of the relationships between different fractal parameters**
**Note**: This figure shows how to recognize the spatial correlation of urban evolution through the radial dimension and the spectral exponent.

**Table 1 The numerical relationships between different fractal parameters**

| Radial dimension, one-point correlation dimension | Point-point correlation dimension, spectral exponent | Self-affine record dimension | Hurst exponent |
|---|---|---|---|

| $D_f$ | $D_0$ | $D_c$ | $\beta$ | $D_s$ | $H$ |
|-------|-------|-------|---------|-------|------|
| 0.5   | 0.5   | -1    | -1      | 3     | -1   |
| 0.75  | 0.75  | -0.5  | -0.5    | 2.75  | -0.75|
| 1     | 1     | 0     | 0       | 2.5   | -0.5 |
| 1.1   | 1.1   | 0.2   | 0.2     | 2.4   | -0.4 |
| 1.2   | 1.2   | 0.4   | 0.4     | 2.3   | -0.3 |
| 1.3   | 1.3   | 0.6   | 0.6     | 2.2   | -0.2 |
| 1.4   | 1.4   | 0.8   | 0.8     | 2.1   | -0.1 |
| 1.5   | 1.5   | 1     | 1       | 2     | 0    |
| 1.6   | 1.6   | 1.2   | 1.2     | 1.9   | 0.1  |
| 1.7   | 1.7   | 1.4   | 1.4     | 1.8   | 0.2  |
| 1.8   | 1.8   | 1.6   | 1.6     | 1.7   | 0.3  |
| 1.9   | 1.9   | 1.8   | 1.8     | 1.6   | 0.4  |
| 2     | 2     | 2     | 2       | 1.5   | 0.5  |
| 2.25  | 2.25  | 2.5   | 2.5     | 1.25  | 0.75 |
| 2.5   | 2.5   | 3     | 3       | 1     | 1    |

**Note**: The scales of numerical value are as follows: $0<D_f, D_0, D_c<2$; $0<\beta<3$; $0<H<1$; $-1< C_\Delta <1$. It is meaningless if the value goes beyond the upper and lower limits.

## 2.4 Spatial correlation and regularity of urban development

At the macro level of cities, spatial correlation analysis of urban growth and form can be made by means of equation (1) and equation (4). The flow chart of spatial correlation analysis based on spectral analysis is displayed in Figure 3. As stated above, equation (1), $\rho(r) = \rho_1 r^{D_f - d}$, is a central correlation function, while equation (4), $C(r) = C_1 r^{2(D_f - d)+1}$, is a density-density correlation function. Besides, at the micro level, a 1-dimensional spatial autocorrelation coefficient associated with the Hurst exponent can be defined in the form (Feder, 1988)

$$C_\Delta = 2^{2H-1} - 1 = 2^{2(d-D_s)-1} - 1, \quad (29)$$

where $d=2$ is the dimension of the embedding space in which urban form is examined. Given different radial dimension values, we can obtain corresponding autocorrelation coefficients or correlation functions by using fractal dimension equations (Table 2).

The results of macro- and micro-correlation analyses show that two radial dimension values are very special: one is $D_f=1.5$, and the other is $D_f=2$. If $D_f=1.5$, the point-point spatial correlation function will equal a constant, i.e., $C(0)=$const. In other words, the correlation function is out of all relations to the distance $r$. If $D_f<1.5$, the point-point spatial correlation function will be in inverse proportion to the distance $r$. This suggests the spatial centripetal force (the strength of concentration) has the advantage over the centrifugal force (the strength of deconcentration). In this case ($D_f$<1.5), the principal way of urban development should be inward space filling. If $D_f$>1.5, the point-point spatial correlation function will be in proportion to the distance $r$. This suggests the spatial centrifugal force has the advantage over centripetal force. In this instance ($D_f$>1.5), the main way of urban development should be outward spatial spread. In theory, the point-point correlation dimension $D_c$ must be less than the one point correlation dimension $D_f$, that is $D_c \leq D_f = D_0$ (Chen, 2011). However, when $D_f$>2, we have $D_c$>$D_f$. This result is unreasonable. This suggests that the radial dimension should be less than 2. If $D_f$<1.5, city growth will be "thin" (urban undergrowth); If $D_f \geq 2$, city growth will be "fat" (urban overgrowth). An inference can be drawn that, if and only if 1.5≤$D_f$<2, urban growth will fit a theoretical fractal growth consistent throughout the scales.

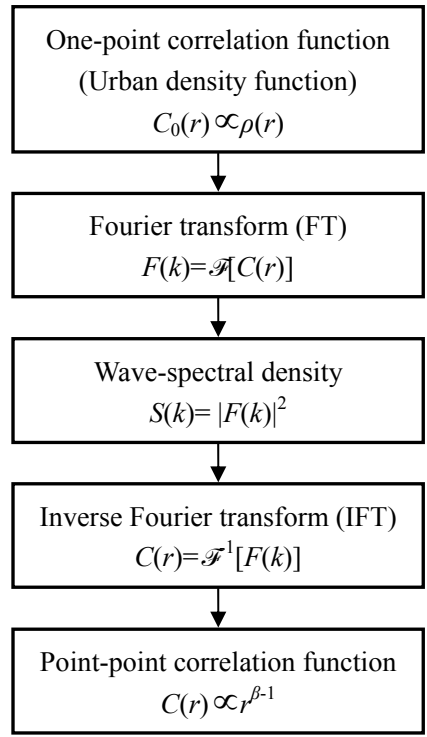

**Figure 3 The procedure of spatial correlation analysis based on spectral analysis**

**Table 2 The autocorrelation coefficients at the micro level and autocorrelation functions at the macro level of cities**

| Radial | Micro level | Macro level |
| --- | --- | --- |

| dimension ($D_f$) | Autocorrelation Coefficient ($C_\Delta$) | One point spatial autocorrelation function [$C_0(r)$] | Point-point spatial autocorrelation function [$C(r)$] |
|---|---|---|---|
| 0.5 | -0.875 | $r^{-1.5}$ | $r^{-2.0}$ |
| 0.75 | -0.823 | $r^{-1.25}$ | $r^{-1.5}$ |
| 1.0 | -0.75 | $r^{-1.0}$ | $r^{-1.0}$ |
| 1.1 | -0.713 | $r^{-0.9}$ | $r^{-0.8}$ |
| 1.2 | -0.670 | $r^{-0.8}$ | $r^{-0.6}$ |
| 1.3 | -0.621 | $r^{-0.7}$ | $r^{-0.4}$ |
| 1.4 | -0.565 | $r^{-0.6}$ | $r^{-0.3}$ |
| **1.5** | **-0.5** | **$r^{-0.5}$** | **$r^0$=Constant** |
| **1.6** | **-0.426** | **$r^{-0.4}$** | **$r^{0.2}$** |
| **1.7** | **-0.340** | **$r^{-0.3}$** | **$r^{0.4}$** |
| **1.8** | **-0.242** | **$r^{-0.2}$** | **$r^{0.6}$** |
| **1.9** | **-0.129** | **$r^{-0.1}$** | **$r^{0.8}$** |
| **2.0** | **0** | **$r^0=1$** | **$r^{1.0}$** |
| (2.25) | 0.414 | ($r^{0.25}$) | ($r^{1.5}$) |
| (2.5) | 1 | ($r^{0.50}$) | ($r^{2.0}$) |

**Note**: The bold numerals represent the proper scale of the fractal parameters of urban form.

Based on the fact that the spectral exponent proved to be a point-point correlation dimension, the theoretical revelation from the spatial correlation analysis of Table 2 can be outlined as below. If $\beta=D_c<1$, i.e. $D_f<1.5$, the spatial correlation intensity varies directly as the distance between two points. This indicates that urban development tends to filling in vacant space (spare land, vacant land) or even open space inwards. In this instance, city planning should be focused on internal space of a city (esp., city proper). If $\beta=D_c>1$, i.e. $D_f>1.5$, the spatial correlation intensity varies inversely (reciprocally) as the distance between two places. This implies urban development tends to growing outwards, and outskirts are gradually occupied by structures, outbuildings, and service areas. In this case, city planning should be focused on external space of a city (suburbs or even exurbs). If $\beta=D_c=1$, i.e. $D_f=1.5$, the point-point spatial correlation intensity is independent of the distance. In this instance, urban evolution seems to be at the self-organized critical state, which takes on inverse power-law distributions (Bak, 1996). The self-organized criticality (SOC) of urban development was discussed by Batty and Xie (1999) and Chen and Zhou (2008).

The radial dimension $D_f=2$ indicates another special value of fractal parameter for urban form. In this case, we have $C_0(0)=1$, this suggests that the one-point spatial correlation is independent of distance. The action of the city center on any urban place is the same as one another. This seems to be inexplicable. In theory, a city center is the location with the highest urban density. If $D_f$=2, the density of one place equals that of another place, and thus the spatial information of city center is covered with various geographical "noises". On the other hand, when $D_f$=2, the autocorrelation coefficient $C_\Delta$=0, and the fBm process will be reduced to Brownian motion (random walk). The space-filling process within the urban area will be ceased owing to absence of vacant space.

In short, different fractal parameters have different ranges of validity (Table 1). The intersection of the valid ranges of $D_0$, $D_c$, $D_s$, $H$, and $\beta$ gives the proper interval or extent of variation of the radial dimension $D_f$, which comes between 1.5 and 2. Corresponding to this range, all the fractal parameters are of meaning and become consistent with each other. If the fractal dimension values go beyond these limits ($D_f$=1.5 and $D_f$=2), the valid relationships between different fractal parameters break down. Despite the scale-free property of fractals, the fractal dimension seems to be a measurement with characteristic scale (length). If the fractal dimension value of a city is too high (say, $D_f$ >2) or too low (say, $D_f$ <1.5), the possible problems of urban development should be investigated. If fractal dimension is too low, the geographical space may be not well developed for a city, or the difference of urban density between the central part and outskirts is too high. By contraries, if fractal dimension is too high, the geographical space may be overly developed, or the difference of urban density between the central part and outskirts is too low. By the way, if the geographical set of points are of utterly uniform distribution, the empirical dimension value will be $D_f$=2, but the theoretical dimension value will be $D_f$=0. This is a result of self-contradiction, and the problem is beyond the scope of discussion in this paper. In the sense of statistical average, the fractal dimension value of urban form is around $D_f$ =1.7 (Batty and Longley, 1994).

## 3 Empirical analysis

### 3.1 Analytical method and steps

Based on the fractal dimension equations and parameter relations, the scaling analysis, spectral analysis, and spatial correlation analysis can be integrated into a new analytical procedure for urban form and growth, and the method can be termed '3S analyses' of cities. The spatial correlation function given above is on the base of continuous variables, but the spatial sampling is a discrete process in practice. Therefore, the correlation functions used for empirical analysis must be calculated on the base of discrete variables. There are two approaches to constructing the density-density correlation functions. One is based on the 1-dimension space, and the other, based on the 2-dimension space. In other words, a point-point correlation function for urban analysis can be defined either in the 1-dimension geographical space or in the 2-dimension geographical space.

Consider a city as a system with $N$ elements. The elements can be abstracted as "particles" or "mass points" on a digital map. If the point-point correlation function is defined in the 1-dimension space, we need a system of concentric circles for spatial sampling; if the correlation function is defined in the 2-dimension space, then we require $N$ sets of concentric circles for spatial measurements (Figure 4). The construction procedure of spatial correlation function based on the 2-dimension space and the related analytical method were illustrated and illuminated by Chen and Jiang (2010). The focus of this paper is only on the point-point correlation defined in the 1-dimension space.

The analytical process of the 1-dimension spatial correlation can be divided into several steps as follows. **Step 1**: Find the centroid or the center of mass of the urban system studied. The city center is always within the central business district (CBD). **Step 2**: Design a set of concentric circles. The interval between any two immediate circles is constant. That is, if we draw a radial from the common center of these circles to the fringe of the system, the points of intersection of the circles are evenly spaced on the line. In theory, the space between two circles can be regarded as a "ring", and the "width" of the rings can be small enough. **Step 3**: Computing average density. This is not only a process of spatial measurement, but also a process of spatial mapping. After the average density of element distribution in each ring is calculated, the geographical information in the 2-dimension space can be mapped into the 1-dimension space. Thus, the rings can be converted into cells, which are distributed along a radial (Figure 1). If the space between two circles is small enough, the size of cells will be small to any degree. **Step 4**: Evaluate the radial dimension by density-radius scaling. In theory, we can estimate the one-point correlation dimension by equation (1), but in practice, it is equation (17) rather than equation (1) that is suitable for us computing the radial dimension. The density series is too sensitive to random noises to give credible results. **Step 5**: Calculate the spectral exponent using wave-spectrum scaling and the fast Fourier transform (FFT). Theoretically, the spectral analysis is by means of equation (9), but empirically, the spectral analysis can be made with equation (15). As mentioned above, the scaling exponent of wave-spectrum relation is just the point-point correlation dimension. **Step 6**: Implement spatial analysis for the urban system. Using the fractal parameter relations shown in Figure 2, we can estimated varied parameter values besides the radial dimension and spectral exponent, including the Hurst exponent and the profile dimension. With the aid of the technical path shown in Figure 3, we can make more sophisticated correlation analysis.

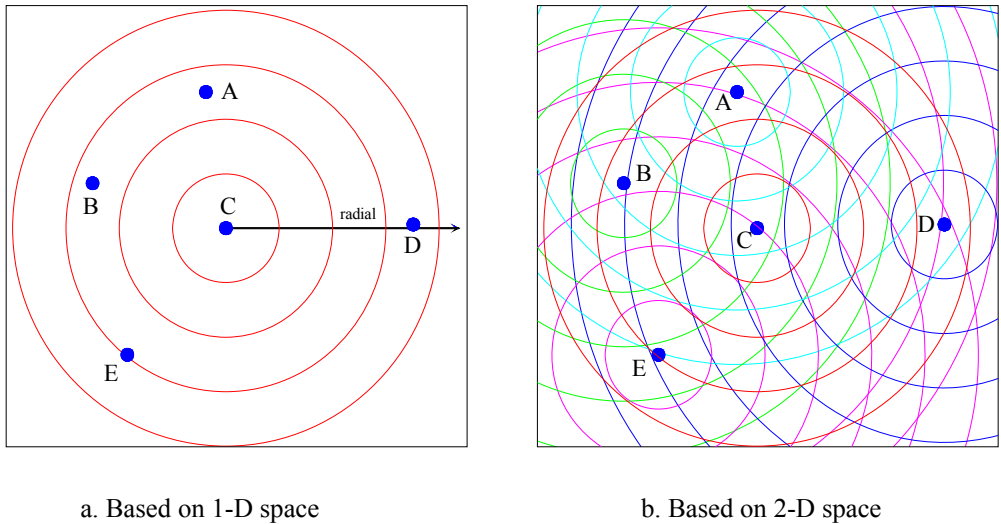

a. Based on 1-D space      b. Based on 2-D space

**Figure 4 The sketch maps of spatial correlation function defined in the 1-dimension space and the 2-dimension space**

**Note:** The patterns based on the simple system of 5 elements are only for reference. For the correlation function defined in the 1-dimension space, we need only one set of concentric circles to calculate average density. For the correlation function defined in the 2-dimension, we need five sets of concentric circles to construct spatial scaling.

### 3.2 Materials and results

The theory and method of the 3S analysis presented above can be applied to the cities in the real world. Three megacities in Yangtze River delta, China, are taken as examples to show how to make use of the fractal parameter equations and the related ideas. The three cities are Shanghai, Nanjing, and Hangzhou (Figure 5). The datasets were extracted from the remotely sensed images (RSIs) in 1985, 1996, and 2005. We have nine datasets of urban land use density for the three cities in three years. As indicated above, there are two approaches to estimating the radial dimension of urban form: the area-radius scaling expressed by equation (17), and the energy-spectrum scaling expressed by equation (9). However, because of the anisotropic growth of urban form, the datasets of the three cities cannot be well fitted to area-radius scaling relations. In this instance, theoretically speaking, we can go by an alternative way and adopt the energy-spectrum scaling relation.

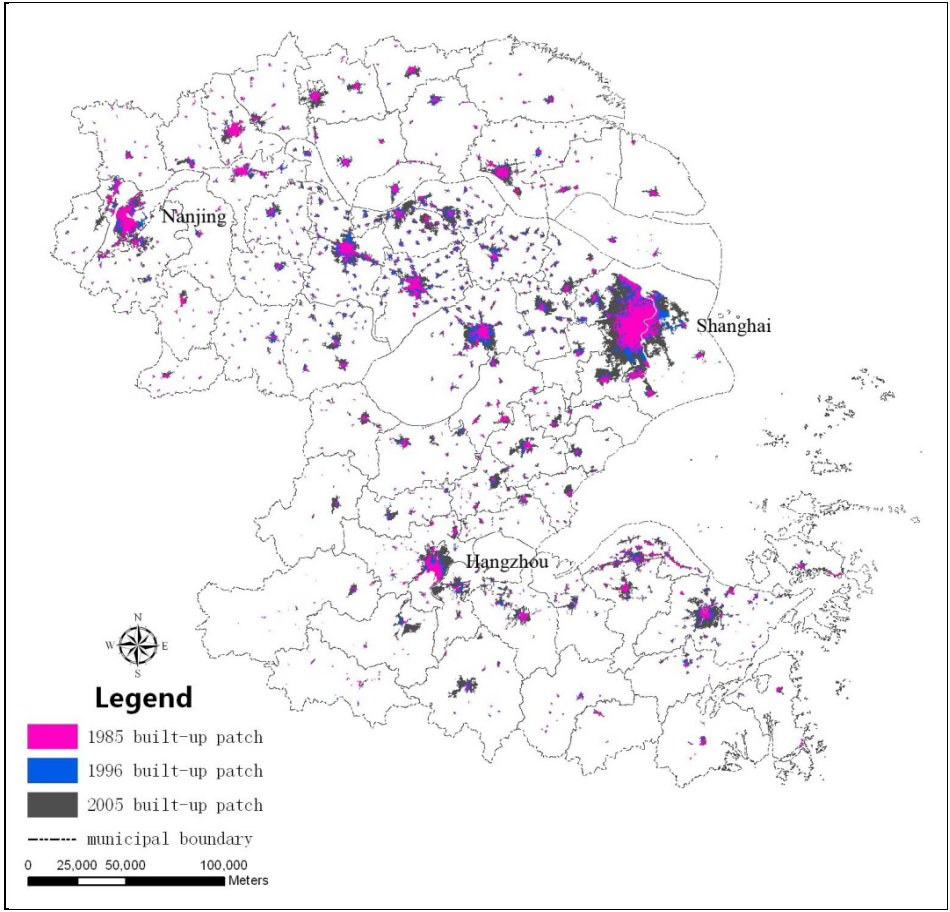

**Figure 5 A sketch map of the main cities in Yangtze River delta, China**

The concept of energy spectrum is based on continuous variable and infinite spatial distance. However, in practice, the urban area is limited and the variables are discrete. Therefore, the energy spectral density, $S(k)$, should be replaced by the wave spectrum density, $W(k)$, which is defined as

$$W(k) = \frac{1}{N} S(k), \qquad (30)$$

where $N$ is the length of the sample path, or the number of data point of urban density. For simplicity, the number is taken as $N=256$ in these cases. Thus, the energy-spectrum scaling is replaced by the wave-spectrum relation such as

$$W(k) \propto k^{-\beta^*}, \tag{31}$$

where $\beta^*$ refers to the estimated value of the spectral exponent. According to equation (10), we can derive a fractal parameter from the wave-spectrum scaling, that is

$$D_f^* = \frac{\beta^*}{2} + 1, \tag{32}$$

which can be termed 'image dimension' (Chen, 2010). A large number of mathematical experiments yielded an empirical formula in the form

$$D_f \approx \frac{2}{5} D_f^* + 1, \tag{33}$$

which reflects the numerical relation between the image dimension and the radial dimension. The image dimension reflects the space filling extent of the central part of a city, while the radial dimension mirror the space filling degree of the whole urban field.

The procedure of data processing and analysis is as follows. **First, compute the average density of urban land use.** Using RSIs, we can determine the city center and estimate the land use density based on the systems of concentric circles. The number of circles is $N$=256. This task can be fulfilled through GIS technique. **Second, calculate the spectral density.** By means of FFT, we can translate the urban density into the spectral density. This task can be fulfilled through the MS Excel or Matlab. **Third, fit the spectral density to the wave-spectrum scaling relation.** The least squares method (LSM) can be employed to do this work. For, example, for Hangzhou in 1985, the wave-spectral relation is as below: $W(k)$=0.0002$k^{-1.8059}$. The goodness of fit is about $R^2$=0.9473, and the spectral exponent is estimated as $\beta^*$=1.8059 (Figure 6). **Fourth, estimate the radial dimension of urban space.** The formula is equation (32). Thus we have $D_f^*$≈1.8059/2+1≈1.9030 for Hangzhou in 1985. The rest may be treated by analogy. By equation (33), the image dimension values can be approximately turned into the radial dimension values. **Fifth, evaluate the related fractal parameters of urban form.** By using the fractal parameter equations, we can compute the spatial correlation dimension, $D_c$, the profile dimension, $D_s$, the Hurst exponent, $H$, and the autocorrelation coefficient, $C_\Delta$, etc. The main results are displayed in Table 4, from which we can find the numerical relationships between different fractal parameters.

Now, the spatio-temporal information of urban evolution of Shanghai, Nanjing, and Hangzhou can be brought to light and the main points are as follows. First, the radial dimension come between 1.75 and 1.77, and the values are very stable from 1985 to 2005. Second, the fractal parameter values of the three cities have no significant differences. This suggests that there existed some coupling relationships between these cities. Third, the radial dimension values are greater than the spatial correlation dimension values, i.e., $D_f > D_c$. This suggests that the spatial structure of the three cities is sound. Fourth, the Hurst exponent is less than 0.5, that is, $H < 1/2$. This implies that the autocorrelation of urban activities are negative, and urban patterns cannot be even and smooth. The positive autocorrelation implies spatial concentration, while the negative autocorrelation suggests that the process of concentration and that of deconcentration are concurrent in spatial distribution of human activities. Generally speaking, the fractal dimension value of urban form goes up gradually with urban growth. Among the three cities, only Shanghai's dimension value went down slightly from 1985 to 2005. This seems to be attributed to the development of the new district of Pudong. It is not the principal aim of this paper to discuss the urban evolution of Yangtze River delta. I will treat them in a companion paper. This section is devoted to showing how to estimate the fractal parameters of real cities.

**Table 4 The fractal parameters of urban form of Shanghai, Nanjing, and Hangzhou in 1985, 1996, and 2005**

| City | Year | $\beta^*$ | $R^2$ | $D_f^*$ | $D_f$ | $D_c$ | $D_s$ | $H$ | $C_\Delta$ |
|---|---|---|---|---|---|---|---|---|---|
| Shanghai | 1985 | 1.7983 | 0.9624 | 1.8992 | 1.7597 | 1.5193 | 1.7403 | 0.2597 | -0.2834 |
|  | 1996 | 1.7965 | 0.9731 | 1.8983 | 1.7593 | 1.5186 | 1.7407 | 0.2593 | -0.2837 |
|  | 2005 | 1.7787 | 0.9694 | 1.8894 | 1.7557 | 1.5115 | 1.7443 | 0.2557 | -0.2872 |
| Nanjing | 1985 | 1.7892 | 0.9355 | 1.8946 | 1.7578 | 1.5157 | 1.7422 | 0.2578 | -0.2852 |
|  | 1996 | 1.8073 | 0.9650 | 1.9037 | 1.7615 | 1.5229 | 1.7385 | 0.2615 | -0.2816 |
|  | 2005 | 1.8102 | 0.9650 | 1.9051 | 1.7620 | 1.5241 | 1.7380 | 0.2620 | -0.2810 |
| Hangzhou | 1985 | 1.8059 | 0.9473 | 1.9030 | 1.7612 | 1.5224 | 1.7388 | 0.2612 | -0.2818 |
|  | 1996 | 1.8134 | 0.9448 | 1.9067 | 1.7627 | 1.5254 | 1.7373 | 0.2627 | -0.2804 |
|  | 2005 | 1.8227 | 0.9816 | 1.9114 | 1.7645 | 1.5291 | 1.7355 | 0.2645 | -0.2785 |

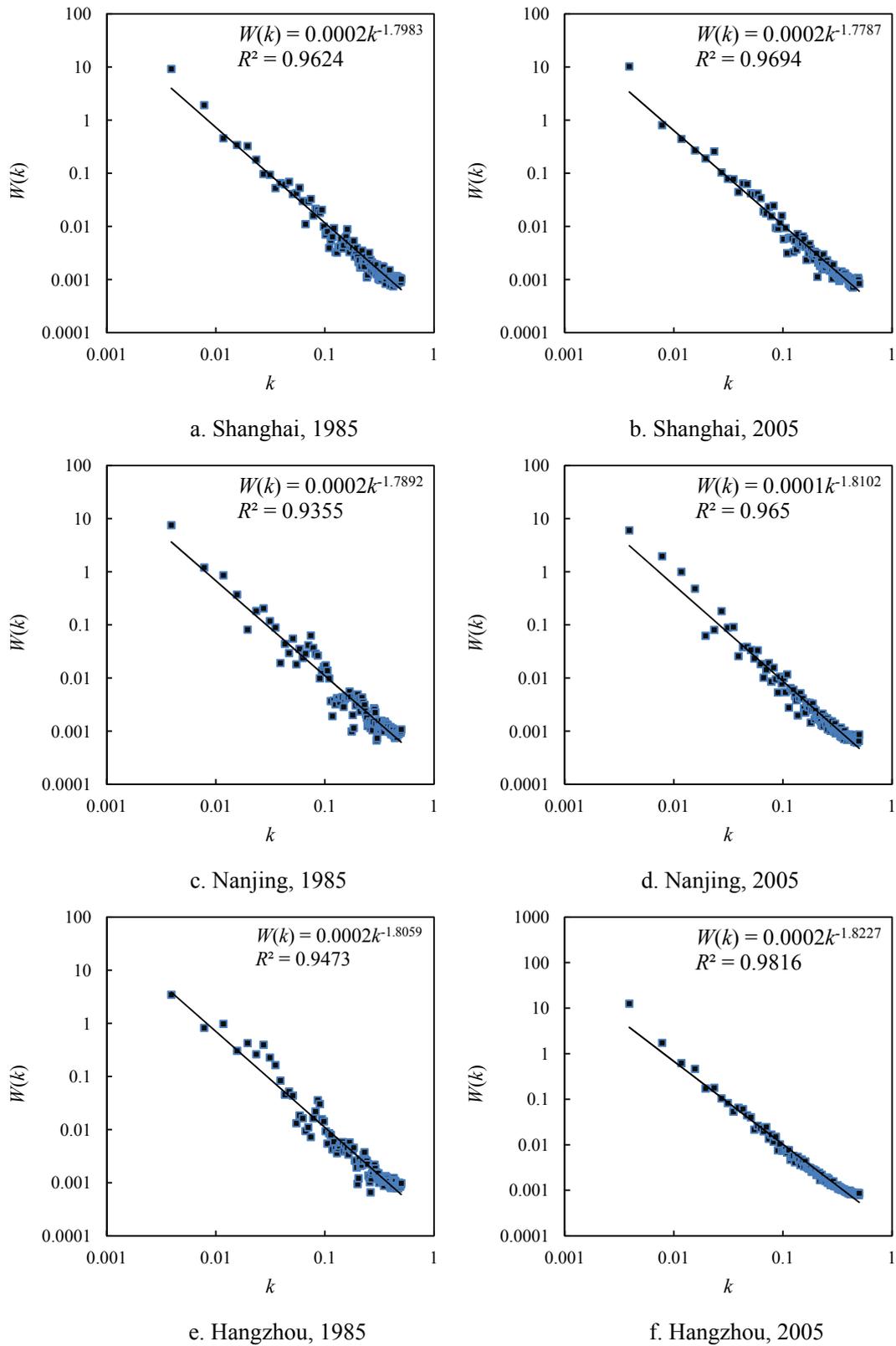

**Figure 6 The wave-spectrum scaling relations of Shanghai, Nanjing, and Hangzhou in 1985 and 2005**

# 4 Discussion

## 4.1 From monocentric cities to polycentric cities

The scaling analysis and spectral analysis of urban form in this paper are based on spatial correlation. The correlation function is defined in the 1-dimension space, but it reflects the geographical information in the 2-dimension space since the fractal dimension values come between 1 and 2, especially for the monocentric cities. Thus, all the fractal parameters associated with the radial dimension indicates the 2-dimension spatial information of cities. If the correlation function is defined in the 2-dimension space, several problems will arise as follows. First, the spatial correlation cannot be directly associated with the urban density function, equation (1), which is very fundamental and simple in urban analysis. Second, it is not easy to make spectral analysis based on Fourier transform, which is very useful for us to reveal geographical regularity and spatial information. Third, the amount of computational work will be very large if the points (elements) are too many. A simple comparison between the two types of correlation functions can be drawn as follows (Table 3). The basic principle of selecting models or methods is the least ratio of "output" (explanation and prediction effects) to "input" (parameter number, workload). In many cases, the output-input ratio of fractal models do not increase if we use the correlation function defined in the 2-dimension space to replace that defined in the 1-dimension space.

Table 3 The similarities and differences between the correlation function defined in the 1-diemsnion space and that defined in the 2-dimension space

| Item | 1-dimension space | 2-dimension space |
| --- | --- | --- |
| Data processing | Statistical average density | Spatial distance matrix |
| Spatial correlation | Regular points in 1-D space | Random points in 2-D space |
| Spectral analysis | Simple process | Complex process |
| Spatial relation | Core-periphery | Network structure |
| Geographical information | 2-dimension space | 2-dimension space |
| Computation work | Smaller workload | Larger workload |
| Suitable object | Cities as systems (urban form) | Systems of cities (city network) |

The function of urban density based on monocentric cities is actually defined in the 1-dimension space. Even if the fractal parameters based on the spatial correlation function for the 1-dimension space reflect the geographical information in a 2-dimension space, they can not reflect ALL the geographical information of urban growth and form. The limitation of the radial dimension and the related parameters are as follows. First, they cannot show the spatial correlation in all directions. The correlational direction is confined to the radial lines from a city center to periphery. Second, they cannot be effectively applied to the spatial form of polycentric cities. For a city with multiple growth centers, the power law of urban density often breaks down. Third, they cannot be used to characterize any components of urban form. Even for the monocentric cities, not all types of urban density follow the inverse power law. The best objects described by Smeed's model are the lines or nodes of traffic network of a city. In many cases, urban population density and urban land-use density data cannot be satisfactorily fitted into the inverse power function.

If we want to investigate the pattern and process of the spatial correlation based on the 2-dimension space, or if the 1-dimension-based correlation function is of no effect because of multicentric growth of cities, two approaches will be available for spatial analyses. One is to adopt the 2-dimension-based correlation function (Chen and Jiang, 2010), and the other is to use the generalized correlation dimension of multifractals (Grassberger, 1985; Hentschel and Procaccia, 1983). The second approach is well known for scientists, but it is hard to comprehend the multifractals concept. Based on the box-counting method, the generalized correlation dimension is always defined in the following form

$$D_q = \frac{1}{q-1} \lim_{\varepsilon \to 0} \frac{\log \sum_{i=1}^{N(\varepsilon)} P_i(\varepsilon)^q}{\log \varepsilon}, \qquad (34)$$

where $D_q$ refers to the order $q$ generalized correlation dimension, $P(\varepsilon)$ to the growing probability of the $i$th fractal copies with linear size $\varepsilon$ in the given level, and $N(\varepsilon)$ to the number of fractal copies in given level. The parameter $q$ is called the order of moment in statistics. If $q=0$, $D_q=D_0$ refers to *capacity dimension*; if $q=1$, $D_q=D_1$ refers to *information dimension*; if $q=2$, $D_q=D_2$ refers to *correlation dimension*. In theory, $q$ is a continuous variable and we have $q \in (-\infty, +\infty)$. However, in practice, $q$ is always a discrete sequence. Generally speaking, we can take an ordered set of quantities such as $q=\ldots,-50, -49, \ldots, -2, -1, 0, 1, 2, \ldots, 49, 50, \ldots$ Thus we get a multifractal spectrum for urban analysis, and this has been discussed in a companion paper. In the empirical studies, the box-counting method can be substituted with the grid methods of fractal dimension estimation (Frankhauser, 1998).

## 4.2 Fractals, emergence, and urban evolution

A fractal is regarded as a pattern or an order emerging from self-organizing evolution of a complex system (Anderson, 1991; Hao, 2004). This is involved with the concepts of *emergence* and *self-organization*. Emergence is central to understanding the level of organization (integrative level) of complex systems such as cities. Emergence is a way of self-organization by which a new pattern or novel structure arises out of interactions of simple elements or relatively simple subsystems (Holland, 1998). Self-organization indicates a spontaneous process of developing in which some form of global order or coordination arises out of a multiplicity of the local interactions between the components of an initially disordered system (Haken and Portugali, 1995; Portugali, 2000). Emergence is actually a special process of self-organization. The property of emergence is a feature of cities (Batty, 2000). It is helpful here to discuss the relationships between fractal structure and emergence of urban evolution.

Fractals are not the inherent structure of urban form, but a result of urban evolution (Benguigui *et al*, 2000). Urban growth is constrained by natural, economic, and social laws and is driven by geographical, historical, and political factors that do not take on evident fractal features. If and only if a city develops to certain stage, a fractal pattern will arise out of spatial interactions of urban components. The appearance of fractal patterns is a process of emergence in city development. Because of emergence, the whole of a city becomes not only more than but also very different from the sum of the parts of the system. Fractal models are useful for us to investigate urban form (a pattern), and the fractal dimension values of different years are helpful for us to understand urban growth (a process). However, the emergence of city fractals implies complex dynamics. The reductionism is inoperative to know how the spatial order comes out of spatial interactions. In order to comprehend urban evolution, we must study the mechanism of emergence in urban self-organization.

The method of 3S analysis and the related parameter relations provide a potential approach to research emergence of urban evolution. The keys for understanding emergence lie in interactions between different parts, the relationship between global level and local level, and the relationship between structure and dynamics of a complex system. The uses of the 3S method are as follows.

First, the 3S analysis can be employed to investigate the spatial interaction of urban components. The spatial correlation functions can be utilized to model the patterns of local interactions of urban parts. Based on spatial correlation, spectral analysis can be used to model the energy distribution of spatial interactions between different elements of a city. Second, the 3S analysis can be employed to explore the relationships between the global and local levels. The scaling analysis is very important for geographers to understand the connection between global and local levels. A scaling process involves varied scales, from the local level to the global level of cities. If a city evolves from a simple state into a complex state, the power-law distribution will emerge, and the scaling analysis can be utilized to reveal the parameter relations and the allometric growth (Chen and Zhou, 2008). The spectral analysis can be applied to revealing the relationships between the spatial correlation at the global level and the autocorrelation at the local level of a city. Third, the 3S analysis can be employed to examine the relationships between dynamics of urban growth and structure of urban form. Because of absence of continuous sampling records within certain period, it is difficult to model the spatial dynamics of city development. However, the spatial structure of a city always contains the dynamic information of urban evolution. The correlation analysis and spectral analysis can be adopted to reveal the spatial information of urban dynamics.

Fractals suggest the optimized structure of physical and human systems. A fractal can fill its space in the most efficient manner. Using the ideas from fractals to plan or design cities will possibly make human beings utilize geographical space in the best way. The preconditions of applying fractal geometry to city planning are as follows. First, the theoretical framework of fractal cities must be constructed. Second, the mechanism of emergence of urban evolution must be revealed. Third, the self-organizing city planning approach must be developed. Maybe one of urgent affairs is to develop the theory of urban self-organization. The self-organizing/organized cities were discussed by Allen (1997), Portugali (2000) and Haken and Portugali (1995), and the relationships between fractal cities and self-organization were discussed by Thomas and Frankhauser *et al* (2008). In particular, the concept of self-organizing planning has been proposed by Portugali (2000).

# 5 Conclusions

The spatial analysis of urban geography is on the threshold of theoretical revolution because of the development of fractal geometry and nonlinear mathematical theory. The traditional concept of scale in geographical analysis will be replaced by the scaling concept, and the traditional distance-based space concept will be substituted with dimension-based space concept (Chen, 2012b). Fractal theory will possibly play an important role in this revolution. This paper is a theoretical and methodological research of fractal cities by using ideas from scaling, symmetry, correlation, and fractal dimension. The fractal parameter relations presented in this work are not only helpful for our developing urban theory, but also useful for future urban management and city planning. The main conclusions of this paper can be summarized as follows.

First, the radial dimension of urban form is the one-point spatial correlation dimension (the zero-order correlation dimension), while the scaling exponent based on the wave-spectral density is the point-point correlation dimension (the second-order correlation dimension). The relation between the one-point correlation and the point-point correlation dimensions can be linked with a simple equation. Using this formula, we can convert the spectral exponent into the radial dimension and *vice versa*. Meanwhile we can convert the one-point correlation dimension into the point-point correlation dimension and *vice versa*. The conversion relations can enhance our comprehension about fractal cities.

Second, a series of fractal parameters of city systems can be associated with one another by the radial dimension and the spectral exponent. The fractal parameters comprise the one-point correlation dimension, the point-point correlation dimension, the Hurst exponent, the self-affine record dimension, and the 1-dimension spatial autocorrelation coefficient. Thus, based on the inverse power law of urban density, scaling analysis, spectral analysis, and spatial correlation analysis can be integrated into a new analytical framework of urban growth and form. The self-similar or self-affine patterns of urban form can be associated with the self-organized process of urban development and evolution.

Third, the reasonable scale of the radial dimension value ranges from 1.5 to 2. If the value of the radial dimension is greater than 2, the relation between the one-point correlation and the point-point correlation will fall into confusion. On the other hand, if the radial dimension value is less than 1, the point-point correlation dimension will become invalid. If the value of the radial dimension is less than 1.5, the relation between the macro level and the micro level of urban structure will be inharmonious. If and only if the radial dimension comes between 1.5 and 2, various fractal parameters of cities will be logical and thus valid meantime, and this suggests that varied relations of urban growth and form become consistent with one another.

**Acknowledgement**

This research was sponsored by the National Natural Science Foundation of China (Grant No. 41171129). The supports are gratefully acknowledged. The author would like to thank Jiejing Wang of The Hong Kong University for providing the essential data on China's urban land use. Many thanks to three anonymous reviewers whose interesting comments were helpful in improving the quality of this paper. One of the reviewers reviewed this paper very carefully and helped me correct a number of clerical errors and improper expressions.

# References

Allen PM. 1997. *Cities and Regions as Self-Organizing Systems: Models of Complexity*. Amsterdam: Gordon and Breach Science

Anderson PW. 1991. Is complexity physics? Is it science? What is it? *Physics Today*, 44 (7): 9-11

Arlinghaus S. 1985. Fractals take a central place. *Geografiska Annaler B*, 67(2): 83-88

Arlinghaus S L, Arlinghaus W C. 1989. The fractal theory of central place geometry: a Diophantine analysis of fractal generators for arbitrary Löschian numbers. *Geographical Analysis*, 21(2): 103-121

Bak P. 1996. *How Nature Works: the Science of Self-organized Criticality.* New York: Springer-Verlag

Batty M. 1995. New ways of looking at cities. *Nature*, 377: 574

Batty M. 2005. *Cities and Complexity: Understanding Cities with Cellular Automata, Agent-Based Models, and Fractals*. London, England: The MIT Press

Batty M. 2008. The size, scale, and shape of cities. *Science*, 319: 769-771

Batty M. 2000. Less is more, more is different: complexity, morphology, cities, and emergence (Editorial). *Environment and Planning B: Planning and Design*, 27(2): 167-168

Batty M, Longley PA. 1994. *Fractal Cities: A Geometry of Form and Function*. London: Academic Press

Batty M, Xie Y. 1999. Self-organized criticality and urban development. *Discrete Dynamics in Nature and Society*, 3(2-3): 109-124

Benguigui L, Czamanski D, Marinov M, Portugali J. 2000. When and where is a city fractal? *Environment and Planning B: Planning and Design*, 27(4): 507–519

Benguigui L, Daoud M. 1991. Is the suburban railway system a fractal? *Geographical Analysis*, 23(4): 362-368

Cauvin C, Reymond H, Hirsch J. 1985. *L'espacement Des Villes: Theorie Des Lieux Centraux Et Analyse Spectrale*. Paris: CNRS Editions

Chen YG. 2008. A wave-spectrum analysis of urban population density: entropy, fractal, and spatial localization. *Discrete Dynamics in Nature and Society*, vol. 2008, Article ID 728420, 22 pages

Chen YG. 2009. Urban gravity model based on cross-correlation function and Fourier analyses of spatio-temporal process. *Chaos, Solitons & Fractals*, 41(2): 603-614

Chen YG. 2010. Exploring fractal parameters of urban growth and form with wave-spectrum analysis. *Discrete Dynamics in Nature and Society*, vol. 2010, Article ID 974917, 20 pages

Chen YG. 2011. Modeling fractal structure of city-size distributions using correlation functions. *PLoS ONE*, 6(9):e24791

Chen YG. 2012a. Fractal dimension evolution and spatial replacement dynamics of urban growth. *Chaos, Solitons & Fractals*, 45 (2): 115-124

Chen YG. 2012b. On the spaces and dimensions of geographical systems. *Journal of Geography and Geology*, 4(1): 118-135

Chen YG, Jiang SG. 2010. Modeling fractal structure of systems of cities using spatial correlation function. *International Journal of Artificial Life Research*, 1(1): 12-34

Chen YG, Zhou YX. 2008. Scaling laws and indications of self-organized criticality in urban systems. *Chaos, Solitons & Fractals*, 35(1): 85-98

De Keersmaecker M-L, Frankhauser P, Thomas I. 2003. Using fractal dimensions for characterizing intra-urban diversity: the example of Brussels. *Geographical Analysis*, 35(4): 310-328

Feder J. 1988. *Fractals*. New York: Plenum Press

Feng J, Chen YG. 2010. Spatiotemporal evolution of urban form and land use structure in Hangzhou,

China: evidence from fractals. *Environment and Planning B: Planning and Design*, 37(5): 838-856

Frankhauser P, Sadler R. 1991. Fractal analysis of agglomerations. In: *Natural Structures: Principles, Strategies, and Models in Architecture and Nature*. Ed. M. Hilliges. Stuttgart: University of. Stuttgart, pp 57-65

Frankhauser P. 1994. *La Fractalité des Structures Urbaines*. Paris: Economica

Frankhauser P. 1998. The fractal approach: A new tool for the spatial analysis of urban agglomerations. *Population: An English Selection*, 10(1): 205-240

Frankhauser P. 2004. Comparing the morphology of urban patterns in Europe - a fractal approach. In: *European Cities – Insights on Outskirts, Report COST Action 10 Urban Civil Engineering, Vol. 2: Structures*. Eds. A. Borsdorf, P. Zembri. Brussels: European Cooperation in the Field of Scientific and Technical Research, pp79-105

Grassberger P. 1985. Generalizations of the Hausdorff dimension of fractal measures. *Physics Letters A*, 107(3): 101-105

Grassberger P, Procaccia I. 1983. Measuring the strangeness of strange attractors. *Physica D*, 9(1-2): 189-208

Haken H, Portugali J. 1995. A synergetic approach to the self-organization of cities and settlements. *Environment and Planning B: Planning and Design*, 22(1): 35- 46

Hao BL. 2004. *Chaos and Fractals*. Shanghai: Shanghai Scientific and Technical Publishers [In Chinese]

Haynes AH. 1975. Dimensional analysis: some applications in human geography. *Geographical Analysis*, 7(1): 51-68

Hentschel HGE, Procaccia I. 1983. The infinite number of generalized dimensions of fractals and strange attractors. *Physica D*, 8(3): 435-444

Holland J H. 1998. *Emergence: from Chaos to Order*. Cambridge, MA: Perseus Books

Hurst HE, Black RP, Simaika YM. 1965. *Long-term Storage: An Experimental Study*. London: Constable

Jiang B, Yao X. Eds. 2010. *Geospatial Analysis and Modeling of Urban Structure and Dynamics*. New York: Springer-Verlag

Makse H, Havlin S, Stanley H E. 1995. Modelling urban growth patterns. *Nature*, 377: 608-612

Makse HA, Andrade Jr. JS, Batty M, Havlin S, Stanley HE. 1998. Modeling urban growth patterns with correlated percolation. *Physical Review E*, 58(6): 7054-7062

Mandelbrot BB. 1983. *The Fractal Geometry of Nature.* New York: W. H. Freeman and Company

Mandelbrot BB. 1989. Fractal geometry: what is it, and what does it do? *Proceedings of the Royal Society of London A: Mathematical and Physical Sciences*, 423 (1864): 3-16

Portugali J.2000. *Self-Organization and the City*. Berlin: Springer-Verlag

Peitgen H-O, Jürgens H, Saupe D. 2004. *Chaos and Fractals: New Frontiers of Science (Second edition)*. New York: Springer-Verlag

Shen G. 2002. Fractal dimension and fractal growth of urbanized areas. *International Journal of Geographical Information Science*, 16(5): 419-437

Smeed RJ (1963). Road development in urban area. *Journal of the Institution of Highway Engineers*, 10(1): 5-30

Takayasu H. 1990. *Fractals in the Physical Sciences*. Manchester: Manchester University Press

Thomas I, Frankhauser P, Biernacki C. 2008. The morphology of built-up landscapes in Wallonia (Belgium): A classification using fractal indices. *Landscape and Urban Planning*, 84(2): 99-115

Thomas I, Frankhauser P, De Keersmaecker M-L. 2007. Fractal dimension versus density of built-up surfaces in the periphery of Brussels. *Papers in Regional Science*, 86(2): 287-308

Thomas I, Frankhauser P, Frenay B, Verleysen M. 2010. Clustering patterns of urban built-up areas with curves of fractal scaling behavior. *Environment and Planning B: Planning and Design*, 37(5): 942-954

Thomas I, Tannier C, Frankhauser P. 2008. Is there a link between fractal dimension and residential environment at a regional level? *Cybergeo: European Journal of Geography*, no.413, 24pages

Wang XS, Liu JY, Zhuang DF, Wang LM. 2005. Spatial-temporal changes of urban spatial morphology in China. *Acta Geographica Sinica*, 60(3): 392-400 [In Chinese]

White R, Engelen G. 1993. Cellular automata and fractal urban form: a cellular modeling approach to the evolution of urban land-use patterns. *Environment and Planning A*, 25(8): 1175-1199

White R, Engelen G. 1994. Urban systems dynamics and cellular automata: fractal structures between order and chaos. *Chaos, Solitons & Fractals*, 4(4): 563-583

# Highlights

- Many fractal parameter relations of cities can be derived by scaling analysis.
- The area-radius scaling of cities suggests a spatial correlation function.
- Spectral analysis can be used to estimate fractal dimension values of urban form.
- The valid range of fractal dimension of urban form comes between 1.5 and 2.
- The traditional scale concept will be replaced by scaling concept in geography.